\documentclass[pre,aps,amssymb,amsmath,superscriptaddress,twocolumn]{revtex4}
\usepackage{amsmath}
\usepackage{amssymb}
\usepackage{epsfig}
\usepackage{amscd}
\usepackage{graphicx,psfrag,xspace}
\usepackage{float}
\usepackage[normalem]{ulem}
\newcommand{\imag}{i}

\newcommand{\glb}{\left(}  
\newcommand{\grb}{\right)}  
\newcommand{\gld}{\left\{}  

\newcommand{\dd}[1]{\text{d}{#1\ }}   

\begin{document}

\title{Fractional Laplacian in Bounded Domains}
\author{A. Zoia}
\email{andrea.zoia@polimi.it}
\affiliation{Department of Physics, Massachusetts Institute of Technology, Cambridge, Massachusetts 02139}
\affiliation{Department of Nuclear Engineering, Polytechnic of Milan, Milan 20133, Italy}
\author{A. Rosso}
\affiliation{Department of Physics, Massachusetts Institute of Technology, Cambridge, Massachusetts 02139}
\affiliation{CNRS - Laboratoire de Physique Th\'eorique et Mod\'eles Statistiques, Universit\'e Paris-Sud, F-91405 Orsay Cedex, France}
\author{M. Kardar}
\affiliation{Department of Physics, Massachusetts Institute of Technology, Cambridge, Massachusetts 02139}
\begin{abstract}
The fractional Laplacian operator, $-(-\triangle)^{\frac{\alpha}{2}}$, appears in a wide class of physical systems, including L\'evy flights and stochastic interfaces. In this paper, we provide a discretized version of this operator which is well suited to deal with boundary conditions on a finite interval. The implementation of boundary conditions is justified by appealing to two physical models, namely hopping particles and elastic springs. The eigenvalues and eigenfunctions in a bounded domain are then obtained numerically for different boundary conditions. Some analytical results concerning the structure of the eigenvalues spectrum are also obtained.
\end{abstract}
\maketitle

\section{Introduction}
\label{Introduction}
Random walks and the associated diffusion equation are at the heart of quantitative descriptions of a large number of physical systems \cite{hughes, feller}. Despite such ubiquity, random walk dynamics has limitations, and does not apply to cases where collective dynamics, extended heterogeneities, and other sources of long-range correlations lead to so-called {\em anomalous dynamics} \cite{klafter, beyondbm, klafter2}. To describe these situations, various generalizations of Brownian motion have been conceived, generally covered under the rubric of {\em fractional dynamics} \cite{klafter}. For example, a quite useful model of {\em super-diffusive} behavior, in which the spread of the distribution grows faster than linearly in time, is provided by L\'evy flights: particles are assumed to perform random jumps with step lengths taken from a distribution that decays as a power law. If the variance of the jump length is infinite, the Central Limit Theorem does not apply \cite{levy, gnedenko, fbm1, levy3, fbm4}, and the dynamics is anomalous. L\'evy flights, which are dominated by rare but etremely large jumps, have proven quite suitable in modeling many physical systems, ranging from turbulent fluids to contaminant transport in fractured rocks, from chaotic dynamics to disordered quantum ensembles \cite{levy1, levy2, levybook, zaslavsky, klafter, klafter2, disorder, disorder2}. 

While the concentration $C(x,t)$ of particles performing Brownian motion follows the standard diffusion equation, ${\partial_t}C(x,t)={\partial_x^2}C(x,t)$, the concentration of L\'evy flights satisfies a {\em fractional diffusion equation} in which the Laplacian operator is replaced by a {\em fractional derivative} as
\begin{equation}
\frac{\partial}{\partial t}C(x,t)=\frac{\partial^\alpha}{\partial |x|^\alpha}C(x,t).
\label{fde}
\end{equation}
In Eq.~(\ref{fde}), $\frac{d^{\alpha}}{d|x|^{\alpha}}$ is the Riesz--Feller derivative of fractional order $\alpha > 0$ \cite{podlubny,samko}, which has an integral representation involving a singular kernel of power-law form (see Appendix \ref{riesz derivatives}). For diffusing particles, the index $\alpha$ roughly characterizes the degree of fractality of the environment, and is in this context restricted to $\alpha \le 2$; for $\alpha > 2$, the correlations decay sufficiently fast for the Central Limit Theorem to hold, and Eq.~(\ref{fde}) is replaced by the regular diffusion equation \cite{feller}.

Interestingly, the same Riesz--Feller derivative also appears in connection with stochastically growing surfaces \cite{majumdar, racz}. In this case, the evolution of the height $h(x,t)$ of the interface is usually written in Langevin form 
\begin{equation}
\frac{\partial}{\partial t}h(x,t)=\frac{\partial^\alpha}{\partial |x|^\alpha}h(x,t)+\eta(x,t),
\label{langevin}
\end{equation}
where $\eta(x,t)$ represents uncorrelated noise of zero mean, and with $\langle\eta(x,t)\eta(x',t')\rangle=2T\delta(x-x')\delta(t-t')$. The fractional derivative mimics the effects of a generalized elastic restoring force. When $\alpha=2$, Eq.~(\ref{langevin}) describes the dynamics of a thermally fluctuating elastic string and is also known as the Edwards-Wilkinson equation \cite{EW}. However, in many physical systems, such as crack propagations \cite{gao} and contact lines of a liquid meniscus \cite{joanny}, the restoring forces acting on $h(x,t)$ are long-ranged and characterized by $\alpha=1$. Other physical systems, such as slowly growing films in Molecular Beam Epitaxy, are better described by a restoring force that depends on curvature, with $\alpha=4$ \cite{toroczkai}.

Better understanding of the properties of the fractional derivative is thus relevant to many physical systems. When the domain over which the operator $\frac{d^{\alpha}}{d|x|^{\alpha}}$ acts is unbounded, the fractional derivative has a simple definition in terms of its Fourier transform
\begin{equation}
\frac{d^{\alpha}}{d|x|^{\alpha}} e^{i q x} = -|q|^{\alpha} e^{i q x}.
\label{def1}
\end{equation}
More precisely, $\frac{d^{\alpha}}{d|x|^{\alpha}}$ is a pseudo-differential operator, whose action on a sufficiently well-behaved function is defined through its symbol $-|q|^{\alpha}$. Another form of the operator, given in Ref. \cite{zaslavsky_def}, is
\begin{equation}
\frac{d^{\alpha}}{d|x|^{\alpha}} : -(-\triangle)^{\frac{\alpha}{2}},
\label{operator}
\end{equation}
where $ (-\triangle) $ is the positive definite operator associated to the regular Laplacian, with symbol $|q|^{2}$. For this reason, $-(-\triangle)^{\frac{\alpha}{2}} $ is also called the {\it fractional Laplacian}. (For $\alpha=2$ we recover the usual Laplacian \cite{podlubny, samko}.)

Thanks to expression~(\ref{def1}), Eqs.~(\ref{fde}) and (\ref{langevin}) on an infinite or periodic support may be easily solved in the transformed space. However, whenever boundary conditions (BC) break translational invariance, Fourier transformation is of limited use, and the long-range spatial correlations (inherent to the non-local nature of the fractional Laplacian operator) make the problem non trivial.

In this paper we investigate the fractional Laplacian on a bounded $1$-$d$ domain with various BC on the two sides of the interval. In particular, we shall study absorbing and free BC: the former naturally arise in the context of L\'evy flights in connection to first-passage problems \cite{buldyrev, levy2}, while the latter arise in the context of long-ranged elastic interfaces with no constraints at the ends \cite{santachiara}. The remainder of the paper is organized as follows: 
in Sec.~\ref{Fractional Laplacian} we recast Eqs.~(\ref{fde}) and (\ref{langevin}) into the eigenvalue problem for the fractional Laplacian. We then introduce a specific discretization of the fractional Laplacian, and present the main advantages of our choice. In Sec.~\ref{boundary conditions} we discuss the implementation of free and absorbing BC by appealing to the examples to L\'evy flights and fluctuating interfaces. The numerical results are presented in Sec.~\ref{numerical results}, with particular emphasis on the behavior of eigenfunctions close to the boundaries. As discussed in Sec.~\ref{analytical results}, some analytical insights into the problem can be achieved by examining certain exactly solvable limits, and by perturbing around them. We end with a concluding Sec.~\ref{conclusions}, and two short appendices. 

\section{Matrix representation of the fractional Laplacian}
\label{Fractional Laplacian}

Consider L\'evy flights in a domain $\Omega \in \cal R$: by applying the standard method of separation of variables, the concentration $C(x,t)$ in Eq.~(\ref{fde}) may be written as
\begin{equation}
C(x,t)=\sum_k \psi_k(x) e^{\lambda_k t} \int_\Omega \psi_k(y) C(y,0) \dd y,
\end{equation}
where $\psi_{k}(x)$ and $\lambda_{k}$ satisfy 
\begin{equation}
 -(-\triangle)^{\frac{\alpha}{2}} \psi_{k}(x) = \lambda_{k}(\alpha) \psi_{k}(x),
\label{problem}
\end{equation}
with the appropriate BC on $\partial \Omega$. Here $-\lambda_{k}$ also corresponds to the inverse of the time constant with which the associated eigenfunction $\psi_{k}(x)$ decays in time. Analogously, in the context of stochastic interfaces, the shape $h(x,t)$ may be decomposed into normal modes $h(x,t)=\sum_k \tilde{h}_k(t) \psi_k(x)$, where $\psi_k(x)$ satisfy Eq.~(\ref{problem}) and $\tilde{h}_k(t)$ are time-dependent coefficients. Substituting this expression for $h(x,t)$ into Eq.~(\ref{langevin}), the normal modes are decoupled from each other, easing the computation of correlation functions.

For the case of an unbounded domain or periodic BC, the set of eigenfunctions and the corresponding spectrum of eigenvalues of the operator in Eq.~(\ref{problem}) is known explicitly \cite{podlubny,samko}. By contrast,  analytical study of Eq.~(\ref{problem}) with different BC is awkward and not completely understood: for absorbing BC it has been proven that the operator $-(-\triangle)^{\frac{\alpha}{2}}$ on a bounded domain admits a discrete spectrum of eigenfunctions and that the corresponding eigenvalues are all real and negative and can be ordered so that $-\lambda_{1}\le-\lambda_{2}\le\cdots \le-\lambda_{\infty}$. However, the exact values of the eigenvalues and the corresponding eigenfunctions are not known and remain an open question (see e.g. Ref. \cite{math} and references therein). It is nonetheless both possible and interesting to investigate the properties of the fractional Laplacian numerically, and at least two major approaches exist for this purpose.

The first approach consists in implementing the continuum operator in Eq.~(\ref{problem}) with a finite differences scheme. This is the so-called Gr\"{u}nwald-Letnikov scheme, whose construction is directly based on the integral representation of the fractional Laplacian operator \cite{gorenflo, gorenflo_scheme, gorenflo_probab}. Considerable insight on the behavior of solutions  to the fractional diffusion equation on unbounded domains is obtained by this method, and it has been shown to be highly accurate. However, due to some technical difficulties, it can not be straightforwardly extended to take into account BC \cite{chechkin,ciesielski,reflecting1}. Another finite element approach to discretization of this continuum operator is presented in Ref. \cite{chen}.

The second approach is intrinsically probabilistic in nature and consists in replacing continuous L\'evy flights representing $\frac{d^{\alpha}}{d|x|^{\alpha}}$ with a discrete hops on a lattice: a transition probability matrix $P_{l,m}$ is constructed, whose elements represent the probability of performing a jump from position $l$ to $m$. Analogous to L\'evy flights, the jump probability has a power-law tail which after normalization reads $P_{l,m} =1/(2\zeta(\alpha+1)|l-m|^{\alpha+1})$, where $\zeta(.)$ is the Riemann Zeta function. For this reason, this process has also been referred to as a Riemann random walk \cite{buldyrev, buldyrev2}. The matrix $D_{l,m}=P_{l,m}-\delta_{l,m}$, is supposed to converge to the representation of the continuum operator when its size goes to infinity. BC can be taken into account by properly setting the probabilities for jumps leading out of the domain. This approach, however, has some shortcomings: first, the convergence of the discretized matrix to the continuum operator largely deteriorates as $\alpha \rightarrow 2$, i.e. when approaching the regular Laplacian \cite{buldyrev,buldyrev2,zoia}. Secondly, it is strictly limited to the range $\alpha \in \left( 0, 2 \right]$, due to its probabilistic underpinnings.

\begin{figure}[t]
   \centerline{ \epsfclipon \epsfxsize=9.0cm
\epsfbox{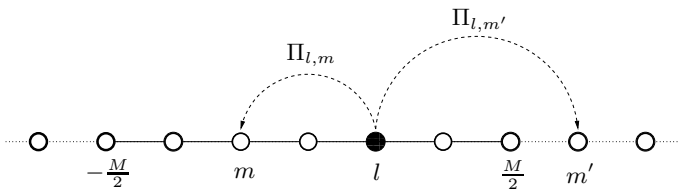} }
   \caption{Implementing BC in a hopping model: for {\em absorbing BC} the jump from $l$ to site $m'$ outside the domain leads to the death of the particle, while for {\em free BC} the jump $(l,m')$ is rejected. For both cases, the jump $(l,m)$ within the interval is accepted.}
   \label{figure_hopping}
\end{figure}

Our approach is the following: we are interested in representing the action of the operator in terms of a matrix $A$ such that the eigenvalues and the eigenvectors of $A$ converge to the eigenvalues and eigenfunctions of the operator when the size $M$ of the matrix goes to infinity. We start with the Fourier representation of the discretized  Laplacian, namely $ -2(1-\cos(q)) $ (in line with the sign convention in Eq.~(\ref{operator})), and raise it to the appropriate power, $- (2(1-\cos(q)))^{\frac{\alpha}{2}}$. The elements of the matrix $A$, representing the fractional Laplacian, are then obtained by inverting the Fourier transform, as
\begin{equation}
A_{l,m} =- \int^{2\pi}_{0} \frac{\dd q}{2\pi} e^{iq(l-m)} \left[ 2(1-\cos(q)) \right]^{\frac{\alpha}{2}}.
\label{integral}
\end{equation}
This is the definition of a {\em Toeplitz} symmetrical matrix $A_{l,m}\left[ \phi \right]$ associated to the generator (the so-called {\em symbol}) $\phi(q) = \left[ 2(1-\cos(q)) \right]^{\frac{\alpha}{2}}$. The generic matrix elements depend only on $n=|l-m|$ and {\em ad hoc} algorithms exist for calculating  the properties of this class of matrices, such as its smallest eigenvalue and the determinant \cite{toeplitz, toeplitz2, toeplitz3}. The integral in Eq.~(\ref{integral}) may be solved explicitly, to give
\begin{equation}
A_{l,m} = A(n) = \frac{\Gamma(-\frac{\alpha}{2}+n)\Gamma(\alpha+1) }{\pi\Gamma(1+\frac{\alpha}{2}+n)} \sin(\frac{\alpha}{2}\pi).
\label{solution}
\end{equation}
In the special cases when ${\alpha}/{2} $ is an integer, $A(n)=(-1)^{\alpha-n+1}C_{\alpha,\frac{\alpha}{2}+n}$, where $C_{\alpha,\frac{\alpha}{2}+n}$ are  binomial coefficients. We remark that $A(n) = 0$ for $n > \alpha/2$, as the poles of $\Gamma(-\frac{\alpha}{2}+n)$ are compensated by the zeros of the $\sin(\alpha\pi/2)$ in Eq.~(\ref{solution}). The off-diagonal elements $ A_{l,m\neq l} $ are all positive when $0<\alpha\leq2$, but come in different signs when $\alpha>2$. Thus, for $\alpha\leq2$ the matrix $A$ can be normalized and interpreted as  transition probabilities for a L\'evy flyer with stability index $\alpha$.

While superficially similar, our approach has notable advantages compared to Riemann walks. The matrix $A$ does not suffer from any deterioration in convergence close to $\alpha=2$, and can in fact be extended beyond the range $0<\alpha\leq2$. The relatively simple structure of the matrix allows to incorporate BC in a straightforward manner. It is also suitable for some analytical treatments, as we will show in detail in the next Sections.

\begin{figure}[t]
   \centerline{ \epsfclipon \epsfxsize=9.0cm
\epsfbox{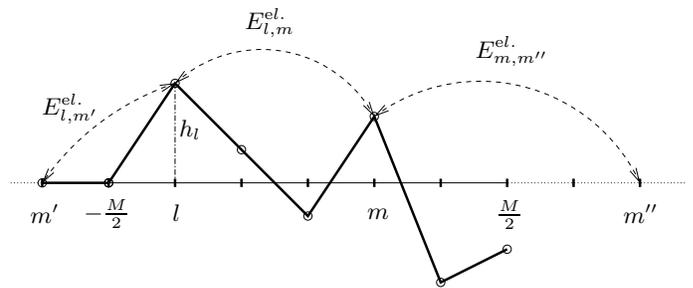} }
   \caption{Implementing BC in a model of elastic springs: {\em Mixed BC} are imposed by removing all springs connected to sites with index $m''>M/2$ ({\em absorbing BC} on the right), and by pinning to zero all sites with index $m'<-M/2$ ({\em free BC} on the left). For the case shown here, $E^{\text el.}_{l,m}= \frac{1}{2} A_{l,m} (h_l-h_m)^2$; $E^{\text el.}_{l,m'}= \frac{1}{2} A_{l,m'} h_l^2$; $E^{\text el.}_{m,m''}=0$. The interface is free to fluctuate at the right boundary and is constrained to zero at the left boundary.}
   \label{figure_spring}
\end{figure}

\section{Boundary conditions for the eigenvalue problem}
\label{boundary conditions}

Due to the non-locality of fractional Laplacian, it is not possible to specify the value of the function $\psi_k(x)$ only locally at the boundaries of a finite domain. Doing so leads to erroneous analytical results, in contrast e.g. with Monte Carlo simulations \cite{dybiec, gitterman, ferraro, dequieroz}. This also implies that standard techniques such as the method of images are not applicable \cite{levy2,chechkin}. Subtle distinctions  which do not appear in the case of regular random walks  need to be introduced, such as between ``first passage" and ``first arrival" times, or between free and reflecting BC \cite{levy2,chechkin}. Therefore, a great amount of ingenuity has been employed  to solve even apparently simple problems such as L\'evy flights constrained to live on the half-axis \cite{zumofen}.

The matrix $A$ introduced in the previous Section is {\it a priori} infinite, thus representing the action of the fractional Laplacian operator on an unbounded domain. Within our approach, BC can be taken into account by modifying the matrix elements related to positions out of the considered domain in a suitable manner, as will be shown in the following. This modification leads in general to a matrix of finite size $M+1$. We will study three different kinds of BC: absorbing on both sides, free on both sides, and mixed (absorbing on the left and free on the right), with reference to two physical models. The first concerns hopping particles, the second elastic springs: both are well defined for $\alpha \le 2$ and absorbing, free and mixed BC are easily implemented. In principle, the set of rules by which we will take into account BC can be extended to an arbitrary $\alpha$.

\subsection{Hopping particles}

Let us consider a particle jumping on a 1-dimensional discrete lattice, as shown in Fig.~\ref{figure_hopping}. When the lattice is infinite, at each time the particle jumps from position $l$ to position $m =l+n$ ($n\ne 0$) with a probability $\Pi_{l,m}=-{A(n)}/{A(0)}$. For $\alpha \le 2$ the probability is well defined if we set $\Pi_{l,l}=0$, as the elements $A_{l \ne m}$ all have the same sign. This model is naturally connected to L\'evy flights, since as shown before $A$ represents the discrete version of the generator of this stochastic process. Let us now discuss how to take into account different BC on an interval $[-M/2,M/2]$.

Absorbing BC are imposed by removing the particle whenever a jump takes it to a site $m$ outside the interval. In the special case of Brownian particles, BC may be assigned locally, since their jumps are of the kind $l \rightarrow l\pm1$ and they must touch the sites $\pm M/2$ in order to leave the interval \cite{feller, levy2, chechkin}. Within our approach, absorbing BC are implemented by cutting the infinite matrix $\Pi$ into a matrix of size $(M+1) \times (M+1)$, thus setting to $0$ all the other elements.

Free BC are implemented as in the Metropolis Monte Carlo approach: if the sampled $m$ lies outside the allowed interval, then the particle is left at its original location $l$. This means that the element $\Pi_{l,l}$ is the probability to stay at $l$. From normalization, clearly we must have $\Pi_{l,l}=1-\sum_{l \ne m}\Pi_{l,m}$. These BC differ from standard reflecting BC as implemented e.g. in Refs. \cite{reflecting1, disorder}, where particles abandoning the interval are sent to their mirror image with respect to the boundary. Free and reflecting BC are identical for Brownian particles, thanks to the locality of jumps.

In the case of mixed BC the particle is removed whenever $m< -M/2$, and remains at $l$ for $m>M/2$. The diagonal element of the matrix thus becomes $\Pi_{l,l}=1/2- \sum_{m=l+1}^{M/2} \Pi_{l,m}$.

\subsection{Elastic springs}

Now consider a network of springs connecting the sites of a 1-dimensional lattice, as shown in Fig.~\ref{figure_spring}.  If the spring constant between sites $l$ and $m$ is $A_{l,m}$, the associated elastic energy is 
\begin{equation}
E^{\text el.} =\sum_{l,m}E^{\text el.}_{l,m}= \sum_{l,m} \frac{1}{2} A_{l,m} (h_l-h_m)^2,
\end{equation}
where $h_l$ is the displacement of site $l$. The elastic force acting on the point $(l,h_l)$, is
\begin{equation}
F(h_l)=-\frac{\delta E}{\delta h_l} = -\sum_{l \ne m} A_{l,m} (h_l-h_m).
\end{equation} 
Such a model also describes the dynamics interfaces with long-range elastic interactions. 
Let us now discuss how to take into account different BC on a bounded interval $[-M/2,M/2]$.

Absorbing BC are implemented in this case by setting $h_m=0$ outside the interval $[-M/2,M/2]$, thus cutting the infinite matrix $A$ into a matrix of size $(M+1)\times(M+1)$. The diagonal elements are now the same as those of the infinite matrix. Physically, this corresponds to fluctuating interfaces pinned to a flat state outside a domain. 

Free BC are implemented by removing all the springs connecting sites inside the interval to sites outside. The diagonal elements of the matrix are then $A_{l,l}=- \sum_{l\ne m} A_{l,m}$. These conditions allow to describe fluctuating interfaces with no constraints at the ends: in the past, these BC have been implemented by using reflecting BC \cite{racz, rosso, ledoussal}. We think that our procedure better represents  the physical situation.

For mixed BC we set $h_m=0$ for $m< -M/2$, and cut all the springs connecting $l$ with $m>M/2$. The diagonal elements of the matrix become $A_{l,l}=A(0)/2 - \sum_{m=l+1}^{M/2} A_{l,m}$.

\begin{figure}[t]
   \centerline{ \epsfclipon \epsfxsize=9.0cm
\epsfbox{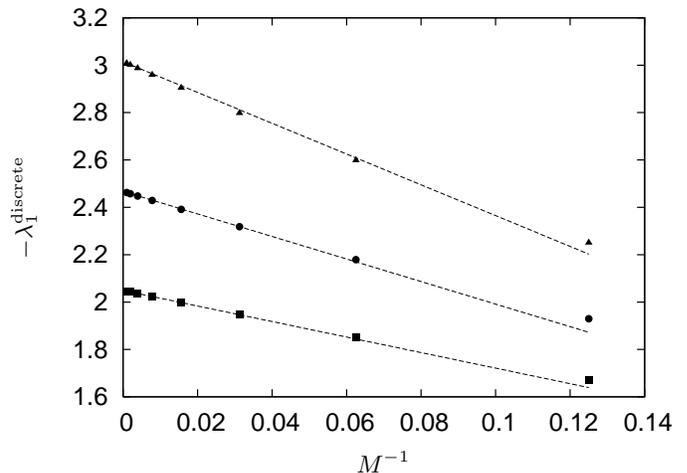} }
   \caption{{\em Absorbing BC:} Convergence of the first eigenvalue with $M$ for $\alpha=1.8, 2, 2.2$. Dashed lines are least-square fits to straight lines, and the continuum limit $\lambda_1(\alpha)$ is obtained for $M^{-1} \rightarrow 0$. }
   \label{slopes_abs}
\end{figure}

\section{Numerical results}
\label{numerical results}

In this Section we discuss our numerical results, as obtained by exploiting the above methods. We will mainly focus on the behavior of the first (non-trivial) eigenfunction of Eq.~(\ref{problem}), which can be regarded as the dominant mode, and of its associated eigenvalue, which represents the inverse of the slowest time constant. For simplicity, in the following we will assume that $\Omega=[-1, 1]$. Given the matrix $A$, which now is modified as to incorporate the appropriate BC, standard numerical algorithms for symmetrical matrices are applied in order to extract the spectrum of eigenvalues and eigenvectors. Then, to obtain the continuum limit, the eigenvalues of $A$ are multiplied by a scale factor $\lambda \rightarrow \lambda (M/L)^\alpha$, where $L=2$ is the size of the interval. We remark that, since the first eigenvalue for free BC is rigorously zero,  we focus on the first non-trivial eigenvalue. The eigenvectors of $A$  are naturally defined only up to a multiplicative factor, and the normalization will be specified later.

\begin{figure}
   \centerline{ \epsfclipon \epsfxsize=9.0cm
\epsfbox{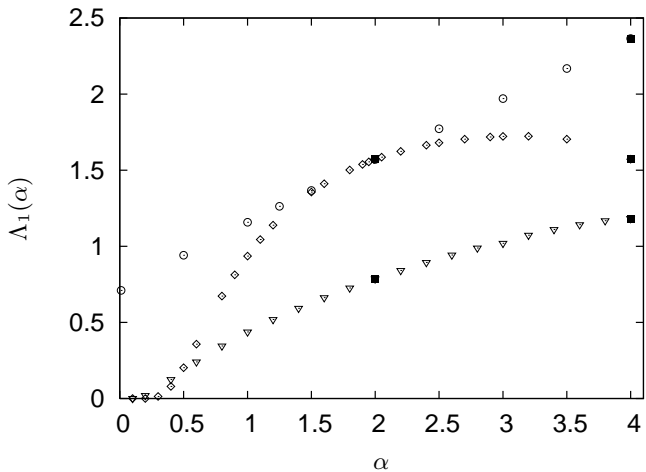} }
   \caption{Eigenvalues with bsorbing (circles), free (diamonds), and mixed (triangles) BC as a function of $\alpha$. Black squares mark the exact values at $\alpha=2$ and $\alpha=4$ (see Section \ref{even}).}
   \label{figure_eigenvalues}
\end{figure}

Let us first discuss the finite-size effects: numerical evidence shows that in the case of absorbing BC the eigenvalues of $A$ converge to the continuum limit $\lambda_{k}(\alpha)$ as $M^{-1}$. The finite-size exponent appears to be exactly $-1$, independent of $\alpha$, while the overall coefficient increases with $\alpha$. These results are depicted in Fig.~\ref{slopes_abs} for the first eigenvalue: the continuum limit is obtained by extrapolating the least-square fit of the convergence plot with $M \rightarrow \infty$. As opposed to Ref. \cite{buldyrev}, our method can be extended to any value of $\alpha$ and does not suffer from any slowing down in convergence as $\alpha\to2$. The extrapolated value for $\alpha=2$ is $\lambda=-2.467\cdots$, extremely close to the expected value of $-\pi^2/4$.

Finite-size effects are very similar for mixed BC, while for free BC the power law convergence for the first non-trivial eigenvalue has an exponent of $-2$ and the slope seems to be approximately constant, independently of $\alpha$.

To explore the structure of the eigenvalues of $A$ for large $M$, i.e. in the continuum limit, let us define
\begin{equation}
\Lambda_k(\alpha)=(-\lambda_k(\alpha))^{\frac{1}{\alpha}}.
\end{equation}
In Fig.~\ref{figure_eigenvalues} we plot the behavior of $\Lambda_k(\alpha)$ as a function of $\alpha$ for absorbing, free, and mixed BC. Note that the eigenvalues of the absorbing BC problem exhibit quite monotonic behavior and actually seem to lie on a straight line: we will come back to this point in Section \ref{even}. Moreover, the eigenvalues of free BC seem to be tangent to those of absorbing BC close to the point $\alpha=2$.

\begin{figure}[t]
   \centerline{ \epsfclipon \epsfxsize=9.0cm
 \epsfbox{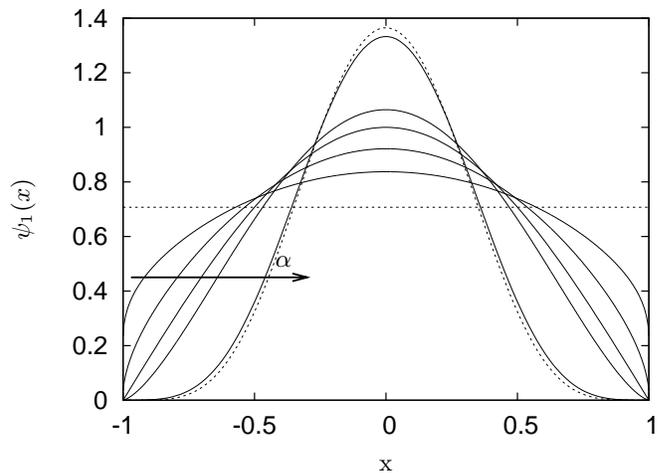}}
   \caption{Eigenfunctions with the smallest eigenvalue $\lambda_1$ for $\alpha=0.1$, 1, 2, 3 and 10 for absorbing BC. The horizontal dashed line corresponds to the limiting function for $\alpha \rightarrow 0$ (marker function). For comparison, we also show for $\alpha=10$ equation Eq.~(\ref{eqn_limit}) as a dotted line.}
   \label{figure_shapes}
\end{figure}

In Fig.~\ref{figure_shapes} we illustrate the shapes of the ground-state eigenfunctions of absorbing BC, corresponding to the first eigenvalue, for different values of $\alpha$. The eigenfunctions have been normalized such that $\int \psi^2_1(x) \dd x=1$. A small and a large value of $\alpha$ have been included to emphasize the limiting behavior at the two extremes: for $\alpha \rightarrow 0$ the eigenfunction seems to converge to the marker function, while for $\alpha \rightarrow \infty$ to a $\delta$ function. It can be shown that the latter limit is approached so that \cite{toeplitz2}
\begin{equation}
\lim_{\alpha \rightarrow \infty} \psi_1(x) = \frac{\Gamma(3/2+\alpha)}{\sqrt{\pi}\Gamma(1+\alpha)} (1-x^2)^{\frac{\alpha}{2}}.
\label{eqn_limit}
\end{equation}

\begin{figure}[t]
   \centerline{ \epsfclipon \epsfxsize=9cm
 \epsfbox{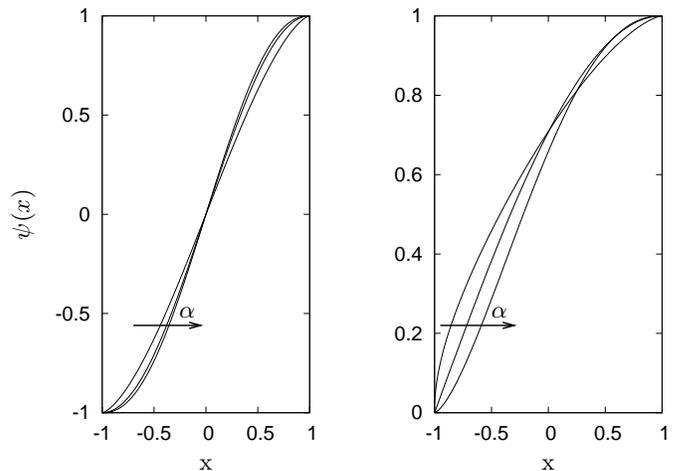}}
   \caption{Eigenfunctions associated with the smallest non-trivial eigenvalue for $\alpha=$ 1, 2, 3, for free (left) and mixed (right) BC.}
   \label{figure_shapes2}
\end{figure}

Typical eigenfunctions for free and mixed BC are depicted in Fig.~\ref{figure_shapes2}. In this case the eigenfunctions have been normalized so that their height ranges respectively in $[-1,1]$ and $[0,1]$.

\begin{figure}[t]
   \centerline{ \epsfclipon \epsfxsize=9.0cm
\epsfbox{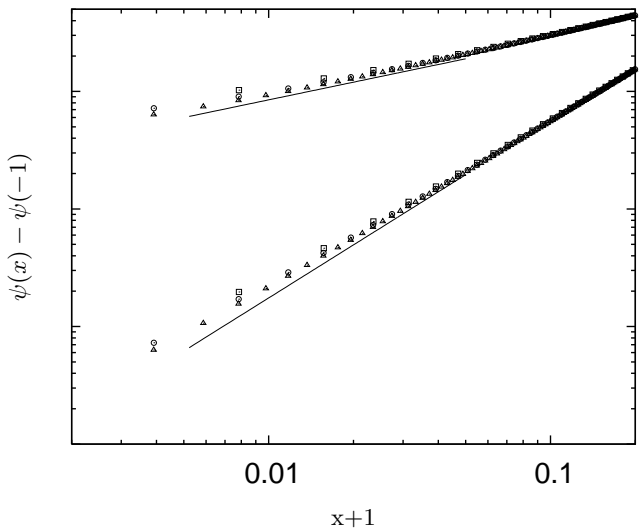} }
   \caption{Scaling of the first eigenfunction close to the boundary for fractional Laplacian of $\alpha=1$, with absorbing (top) and free (bottom) BC. Symbols correspond to numerical eigenvectors for $M=256$, 512, 1024, while solid lines correspond to $(x+1)^{1/2}$ and $(x+1)^{3/2}$, respectively.}
   \label{figure_limit1}
\end{figure}

An important question is how eigenfunctions behave close to the boundaries. As a specific case, we focused on the case $\alpha=1$, and for absorbing BC, our numerical results indicate $\psi_1(x) \sim (1-|x|)^{1/2}$ as $x \rightarrow \pm 1$ (see Fig.~\ref{figure_limit1}). This result is consistent with the findings of Refs. \cite{zumofen, buldyrev2}, which show that in general for absorbing BC the eigenfunctions scale as $(-|x|+1)^{\alpha/2}$. The limiting behavior for free BC in Fig.~\ref{figure_limit1} is less clear: the convergence is rather poor, and we are unable fully characterize the dependence of the slope on $\alpha$. Nonetheless, we can exclude the simplest ansatz that the eigenfunction for a generic $\alpha$ scales linearly close to the boundaries, as suggested by the behavior at $\alpha=2$ and $\alpha=0$, where $\psi_1(x) \sim (1-|x|)^{1}$. In fact, the fit in Fig.~\ref{figure_limit1} is for an exponent $\alpha/2+1=3/2$.

\section{Analytical results for absorbing BC}
\label{analytical results}

For the case of absorbing BC it is possible to derive further information on the structure of the eigenvalues of Eq.~(\ref{problem}) by resorting to analytical treatment.
 
\subsection{Even $\alpha$, and general structure of the eigenvalues}
\label{even}

When $\alpha$ is an even integer, the eigenvalue-eigenfunction Eq.~(\ref{problem}) may be cast in a different way. In particular, Eq.~(\ref{def1}) can be extended to complex $q$ by omitting the absolute value. Then, since $\lambda = -q^{\alpha}$ is real and negative, we can associate to each $\lambda_k$, $\alpha$ independent solutions characterized by $q_j = \Lambda_k \omega_j$, for $j=0,~1,\cdots,\alpha-1$, where $\omega_j=\cos(2\pi j/\alpha)+\imag \sin(2\pi j/\alpha)$ are the $\alpha$ roots of unity. The general form of an eigenfunction is
\begin{equation}
\psi_k(x)= \sum_{j=0}^{\alpha-1} c_{j,k} e^{i \Lambda_k \omega_j x},
\label{general_eigenvector}
\end{equation}
where $c_{j,k}$ are to be determined by imposing the BC
\begin{equation}
\psi_k( \pm 1)= \psi_k^{(1)}(\pm 1)=\psi_k^{(\alpha/2-1)}(\pm 1)=0.
\end{equation}
Thus, determining $\Lambda_k$ is equivalent to finding the zeros of the determinant of the $\alpha \times \alpha$ matrix $B$
\begin{equation}
B =
   \glb \begin{array}{cccc} 
  e^{ i \Lambda \omega_0} & \cdots  & e^{ i \Lambda \omega_{\alpha-1}} \\
  e^{- i \Lambda \omega_0} & \cdots  & e^{ -i \Lambda \omega_{\alpha-1}} \\
   \vdots & & \vdots  \\
  \omega_0^{\alpha/2-1}e^{ i \Lambda \omega_0} & \cdots  &  \omega_{\alpha-1}^{\alpha/2-1}e^{ i \Lambda \omega_{\alpha-1}} \\
  \omega_0^{\alpha/2-1}e^{ -i \Lambda \omega_0} & \cdots  &  \omega_{\alpha-1}^{\alpha/2-1}e^{ -i \Lambda \omega_{\alpha-1}} \\
   \end{array} \grb.
   \label{B}
\end{equation}

The structure of the function $\det(B)=0$ is rather involved. However, for large $k$ it is possible to rewrite this equation in the following form
\begin{equation}
f_\alpha(\Lambda_k) \cos(2 \Lambda_k) + g_\alpha( \Lambda_k) =0,
\end{equation}
when $\alpha/2$ is even and
\begin{equation}
f_\alpha(\Lambda_k) \sin(2 \Lambda_k) + g_\alpha( \Lambda_k) =0,
\end{equation}
when $\alpha/2$ is odd. Here $f_\alpha(\Lambda_k)=\cosh(2 \cot (\pi/\alpha) \Lambda_k)$ and 
\begin{equation}
\frac{g_\alpha(\Lambda_k)}{f_\alpha(\Lambda_k)} \sim e^{-2 \sin(\frac{2 \pi}{\alpha}) \Lambda_k},
\end{equation}
when $k \rightarrow \infty$.

Two special cases need to be considered separately: for $\alpha=2$ we have $g_2( \Lambda_k) =0$ and for $\alpha=6$ an acciddental factorization gives $g_6( \Lambda_k) = \sin(\Lambda_k) \left(\cosh(\sqrt{3} \Lambda_k) + \cdots \right)$. This allows to conclude that for large $k$ the roots of $\det(B)=0$ converge exponentially fast to those of $\cos(2 \Lambda_k)=0$ when $\alpha/2$ is even or $\sin(2 \Lambda_k)=0$ when $\alpha/2$ is odd. These asymptotic roots are exact for $\alpha=2$ for every $k$ and for $\alpha=6$ for all odd $k$, thanks to the factorization.

\begin{figure}
   \centerline{\epsfclipon \epsfxsize=9.0cm
\epsfbox{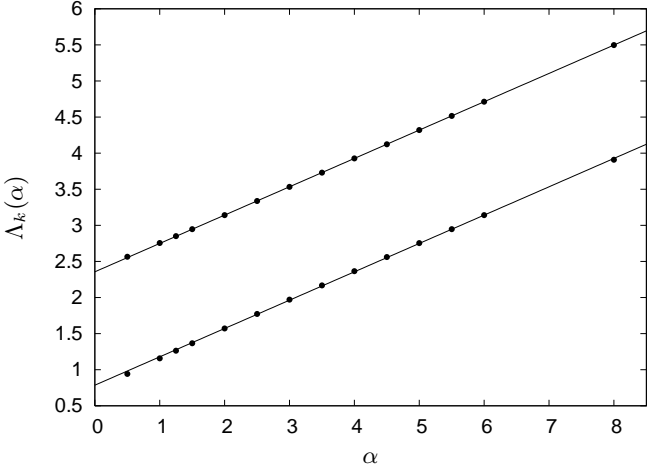} }
   \caption{$\Lambda_k$ as a function of $\alpha$ for $k=1$ and $2$ (dots), compared to the approximation in  Eq.~(\ref{straight_linek}) (straight lines).} \label{spectrum}
\end{figure}

\begin{figure}
   \centerline{ \epsfclipon \epsfxsize=9.0cm
   \epsfbox{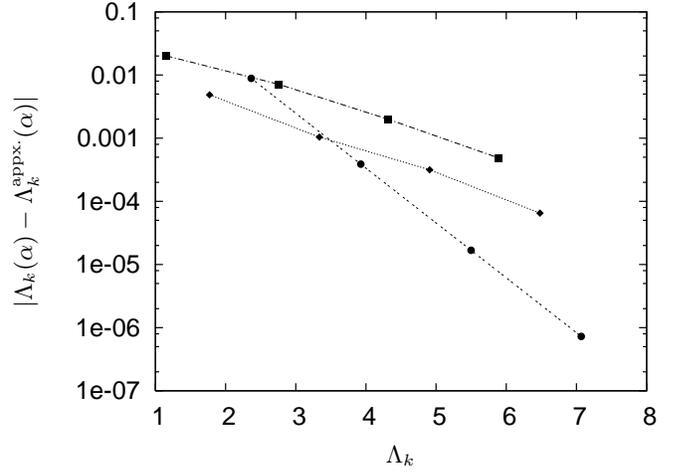} } 
\caption{The difference between $\Lambda_k(\alpha)$ to $\Lambda_k^{\text{appx.}}(\alpha)$ for $\alpha=1$ (squares), $\alpha=2.5$ (diamonds) and $\alpha=4$ (dots), as a function of $\Lambda_k$.}
\label{convergence}
\end{figure}

These considerations, together with the fact that $\Lambda_k(\alpha)<\Lambda_k(\alpha+2)$, allow to state that the eigenvalues $\Lambda_k(\alpha)$ as a function of $k$ will be better and better described by a monotonically increasing function whose simplest form is the straight line
\begin{equation}
\Lambda_k^{\text{appx.}}(\alpha)=\frac{\pi}{8} \alpha +\frac{\pi}{4}(2k-1).
\label{straight_linek}
\end{equation}
Equation~(\ref{straight_linek}) is consistent with our numerical findings and generalizes an observation by Rayleigh,  that for $\alpha=4$ the two values $\Lambda_k(\alpha)$ and $\Lambda_k^{\text{appx.}}(\alpha)$ are identical to the sixth decimal digit for $k \ge 4$ \cite{rayleigh}. In particular we remark that direct numerical evaluation of $\det(B)=0$ reveals that Eq.~(\ref{straight_linek}) is a very good approximation even for $k=1$ if $\alpha$ is not too large, while it has been shown that for very large $\alpha$ the asymptotic behavior of the first eigenvalue is \cite{toeplitz2}
\begin{equation}
\Lambda_1(\alpha)=(4 \alpha \pi)^{\frac{1}{2\alpha}}\frac{\alpha}{e}.
\label{correction}
\end{equation}
Surprisingly, the asymptotic form of Eq.~(\ref{straight_linek}) is valid also for a generic real $\alpha$, as shown in Fig.~\ref{spectrum} for $k=1$ and $k=2$. Setting aside some special cases of $\alpha$ such as $2$ and $4$, to our best knowledge the approximation in Eq.~(\ref{straight_linek}) is a new result. To illustrate the trends, the error in the approximation in depicted in  Fig.~\ref{convergence}. In all cases considered, numerical results indicate that the error vanishes exponentially for large $k$, in agreement with the analytical findings for even $\alpha$.

\subsection{Perturbation theory}

We next examine the behavior of eigenvalues close to $\alpha=2$ and $\alpha=0$ using standard perturbation theory. Throughout this Section we will consider a symmetric domain $\Omega=[-L/2,L/2]$. \linebreak 

\subsubsection{Perturbation around $\alpha=2$}

The ground state eigenvector for $\alpha=2$ on the discrete interval $[-M/2,M/2]$ is 
\begin{equation}
\psi_1(l) = \sqrt{\frac{2}{M}} \cos\left(\frac{\pi l}{M}\right),
\end{equation}
with a corresponding eigenvalue of
\begin{equation}
\lambda_1 = \left(\frac{M}{L}\right)^\alpha \langle \psi_1|A| \psi_1 \rangle,
\end{equation}
where $L$ is the length of the interval. In order to deal with dimensionless quantities, we multiply $\lambda_1$ by $L^\alpha$, and set
\begin{equation}
\hat{\lambda}_1 = \lambda_1 L^\alpha = M^\alpha \langle \psi_1 |A| \psi_1 \rangle.
\end{equation}
For $\alpha=2$, where $A(0)=-2$, $A(1)=1$ and $A(n>1)=0$, we have
\begin{equation}
\hat{\lambda}_1= -M^2 \left[2-2\cos(\frac{\pi}{M}) \right] \sim -\pi^2.
\end{equation}
Setting $\alpha=2+\epsilon$, the operator $A(n)$ becomes, at the first order in $\epsilon$:
\begin{equation}
A(n)=
\gld
\begin{array}{cc}
-2-\epsilon & \text{ for } n=0\\
1+\frac{3}{4}\epsilon & \text{ for } n=1\\
-\frac{1}{(n+1)n(n-1)}\epsilon & \text{ for } n>1
\end{array}
\right. .
\end{equation}

The correction to the ground state is given by
\begin{equation}
\hat{\lambda}_1^\ast =  \hat{\lambda}_1 + \delta \hat{\lambda} = M^{2+\epsilon} \langle \psi_1 |A| \psi_1\rangle,
\label{perturbedeqn}
\end{equation}
which can be rewritten in the following way:
\begin{equation*}
\frac{\hat{\lambda}_1^\ast}{M^{2+\epsilon}} =A(0) +2\sum_{n=1}^{M} A(n) \sum_{l=-M/2}^{M/2-n} \psi_1(l)\psi_1(l +n).
\end{equation*}
By noticing that
\begin{equation*}
\sum_{l=-M/2}^{M/2-n} \psi_1(l)\psi_1(l +n)=
\frac{M-n}{M}\cos\left(\frac{n\pi}{M}\right)+\frac{1}{\pi}\sin\left(\frac{n\pi}{M}\right),
\end{equation*}
we can rewrite the previous expression as
\begin{equation*}
\hat{\lambda}_1^\ast =- M^{2+\epsilon} \left( \frac{\pi ^2}{M^2} + \epsilon Q \right),
\end{equation*}
where $Q$, in the limit of large $M$, is given by
\begin{multline*}
Q = -\frac{1}{2} + \frac{3}{4}\frac{\pi^2}{M^2} + 2 \sum_{n=2}^{M} A(n)\left(1-\frac{1}{2}\frac{n^2\pi^2}{M^2}\right)\\
+\frac{2}{M^2}\int_0^1 \dd x\frac{(1-x)\cos(\pi x)+\frac{\sin(\pi x)}{\pi}-1+\frac{\pi^2 x^2}{2}}{x^3}.
\end{multline*}
Performing the integration, we find
\begin{equation*}
Q M^2 = -\pi^2 \log(M)+\pi\left(\text{Si}(\pi)+\pi\log(\pi)-\pi\text{Ci}(\pi)\right),
\end{equation*}
where $\text{Si}$ and $\text{Ci}$ are the Integral Sine and Integral Cosine functions, respectively.
We can finally come back to $\lambda_1^\ast$, which, expanding for small $\epsilon$, reads
\begin{equation}
\hat{\lambda}_1^\ast = -\pi^2 + \epsilon \left[\pi^2\text{Ci}(\pi)-\pi\text{Si}(\pi)-\pi^2\log(\pi) \right].
\end{equation}
This approach can be extended to eigenfunctions $\psi_k(l)$ of every order $k$. By replacing $\psi_1(l)$ into Eq.~(\ref{perturbedeqn}) with the generic $\psi_k(l)$ (see Appendix \ref{regular laplacian}) and performing the summations as shown above, after some algebra we find the first-order correction  $\delta \hat{\lambda}_k=\hat{\lambda}_k^\ast -\hat{\lambda}_k $, with
\begin{equation}
\delta \hat{\lambda}_k=\epsilon \left[k^2\pi^2\text{Ci}(k\pi)-k\pi\text{Si}(k\pi)-k^2\pi^2\log(k\pi) \right].
\label{expr_pert}
\end{equation}

\begin{figure}
   \centerline{ \epsfclipon \epsfxsize=9.0cm
   \epsfbox{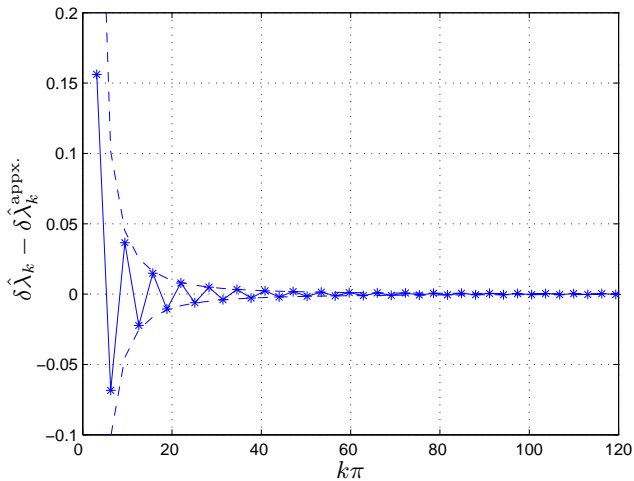} } 
\caption{The error in slope of $\delta \hat{\lambda}_k$, compared to Eq.~(\ref{expr_analy}) for $\alpha=2$ as a function of $k\pi$ (asterisks). The enveloping dashed curves are $\pm4/(k\pi)^2$.}
\label{perturbation_k}
\end{figure}

Now,  consider the curve $\lambda_k^{\text{appx.}}$, which after rescaling by a factor $L^\alpha$ gives
\begin{equation}
\hat{\lambda}_k^{\text{appx.}}=-\left[ \frac{\pi}{4} \alpha +\frac{\pi}{2}(2k-1) \right]^\alpha.
\end{equation}
By putting $\alpha \rightarrow 2 + \epsilon$ and expanding for small $\epsilon$, we get 
\begin{equation}
\delta \hat{\lambda}_k^{\text{appx.}}=\epsilon \left[-k \frac{\pi^2}{2} - k^2\pi^2 \log(k\pi) \right].
\label{expr_analy}
\end{equation}
We can thus compare Eq.~(\ref{expr_pert}), which derives from the perturbative calculations, with Eq.~(\ref{expr_analy}), which stems from our  generic approximation to the eigenvalues of Eq.~(\ref{problem}). In Fig.~\ref{perturbation_k} we plot the error $\delta \hat{\lambda}_k-\delta \hat{\lambda}_k^{\text{appx.}}$ as a function of $k \pi$. As $k$ increases, the slope of the curve along which the actual eigenvalues lie in the proximity of $\alpha=2$ approaches very rapidly to the slope of the curve $\hat{\lambda}_k^{\text{appx.}}$.

We have also applied perturbation theory for $\alpha=2$ to the case of free BC, for which the eigenfunctions are known analytically (see \ref{regular laplacian}). Calculations analogous to those leading to Eq.~(\ref{expr_pert}) allow to derive $\delta \hat{\lambda}_k$ as
\begin{multline}
\delta \hat{\lambda}_k=\epsilon \left[4+ k^2\pi^2\text{Ci}(k\pi)+\right.\\
\left.-3k\pi\text{Si}(k\pi)-k^2\pi^2\log(k\pi)+2k\pi\text{Si}(2k\pi) \right].
\end{multline}
The values of $\delta \hat{\lambda}_k$ for free BC are close but not equal to those of absorbing BC, thus ruling out the hypothesis that the curves $\Lambda_k(\alpha)$ for free and absorbing BC are tangent near the point $\alpha=2$. \linebreak 

\subsubsection{Perturbation around $\alpha=0$}

When $\alpha$ is $0$, $\frac{d^{0}}{d|x|^{0}}$ becomes the identity operator $-I$ and the associated first (and only) eigenvalue is $\lambda_1(\alpha)=1$. In principle, for $\alpha=0$ the operator is highly degenerate, but considering the limiting behavior and the scaling behavior near the boundaries we are led to conclude that the discrete ground-state eigenvector for $\alpha=0$ is
\begin{equation}
\psi_1(l) =\frac{1}{\sqrt{M+1}} I_\Omega(l),
\end{equation}
where $I_\Omega(l)$ is the marker function of the domain $\Omega=[-M/2,M/2]$ (see Fig.~\ref{figure_shapes}). Setting $\alpha=0+\epsilon$, the operator $A(n)$ is corrected at the first order as
\begin{equation}
A(n)=
\gld
\begin{array}{cc}
-1 + o(\epsilon^2) & \text{ for } n=0\\
\frac{1}{2n} \epsilon & \text{ for } n>0
\end{array}.
\right.
\end{equation}
The correction to the ground state is given by
\begin{equation}
\hat{\lambda}_1^\ast = \frac{M^{\epsilon}}{M+1} \sum_{l,m} I_{\Omega}(l) A(n) I_{\Omega}(m),
\end{equation}
which in the limit of large $M$ is
\begin{equation}
\hat{\lambda}_1^\ast = -M^{\epsilon} \left[1 - \epsilon \log(M) +\epsilon (1-\gamma) \right],
\end{equation}
where $\gamma=0.57721566\cdots $ is the Euler-Mascheroni constant. Expanding for small $\epsilon$, we finally get
\begin{equation}
\hat{\lambda}_1^\ast = -1 - \epsilon \left(1-\gamma\right).
\end{equation}
This value is to be compared with $\hat{\lambda}_1^{\text{appx.}}$, which for $\alpha=0+\epsilon$ reads
\begin{equation}
\hat{\lambda}_1^{\text{appx.}}=-1-\epsilon \log\left(\frac{\pi}{2}\right).
\end{equation}

\subsection{First passage time distribution}
\label{first passage}

Knowledge of the fractional Laplacian operator allows us to address the temporal behavior of the L\'evy flyer concentration $C(x,t|x_0)$, where $x_0$ is the starting position of walkers at $t=0$. For example, let us consider the first passage time distribution for the one-dimensional bounded domain $\Omega$ with absorbing BC on both sides, which is obtained as \cite{redner}:
\begin{equation}
\rho(t|x_0)=- \frac{\partial}{\partial t} \int_{\Omega} \dd x C(x,t|x_0).
\end{equation}
In particular, moments of the distribution $\rho(t|x_0)$ are given by
\begin{multline}
\langle t^m \rangle(x_0)=\int_0^\infty  \dd t t^m \rho(t|x_0) =\\
= -\int_0^\infty  \dd t t^m \frac{\partial}{\partial t} \int_{\Omega} C(x,t|x_0) .
\end{multline}
For $m=1$, integrating by parts a using the relation
\begin{equation}
\frac{\partial}{\partial t}C(x,t|x_0)=\frac{\partial ^\alpha}{\partial |x_0|^\alpha}C(x,t|x_0),
\end{equation}
we get
\begin{multline}
\frac{\partial ^\alpha}{\partial |x_0|^\alpha}\langle t^1 \rangle(x_0) =\\
=\int_{\Omega} \dd x C(x,\infty|x_0) -\int_{\Omega} \dd x C(x,0|x_0) =-1.
\end{multline}

\begin{figure}[t]
   \centerline{ \epsfclipon \epsfxsize=9.0cm 
\epsfbox{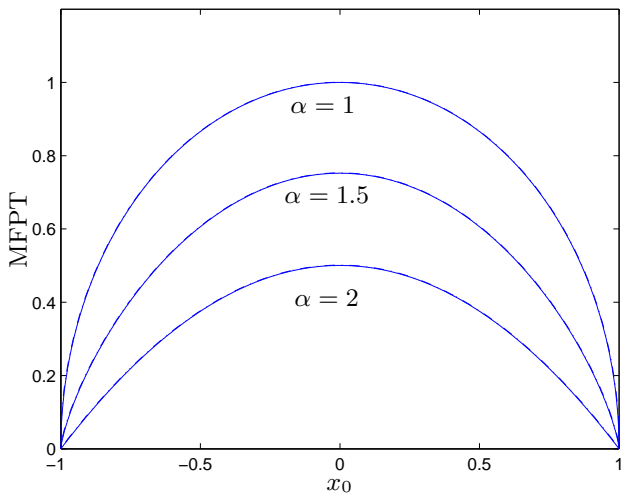} }
   \caption{MFPT as a function of the starting point $x_0$ for $\alpha=1$, 1.5 and 2. Here $L=2$ and $M=1024$. Solid lines are the analytical result $\langle t^1 \rangle(x_0)=(1-x^2_0)^{\alpha/2}/\Gamma(\alpha+1)$, while dashed lines are obtained from the numerical solution $\langle t^1 \rangle(x_0)=-A^{-1}{\bf 1}(2/M)^{\alpha}$. In the limit of large $M$, the two results are in complete agreement for all $x_0$ and $\alpha$.}
   \label{mfpt}
\end{figure}

This equation for the mean first passage time (MFPT) may be solved analytically in closed form (see Ref. \cite{buldyrev2}, and references therein), to give $\langle t^1 \rangle(x_0)=((L/2)^2-x^2_0)^{\alpha/2}/\Gamma(\alpha+1)$, where $L$ is the length of the bounded interval (we have assumed that the interval is symmetric around the origin $x=0$). In Fig.~\ref{mfpt} we compare this expression with the numerical solution obtained by replacing the fractional Laplacian with the discrete operator $A$, namely $ \langle t^1 \rangle(x_0)=-A^{-1}{\bf 1}(L/M)^{\alpha}$; the two curves are in excellent agreement for all $\alpha$ and $x_0$. We remark that the required inversion of the discrete operator may be efficiently performed thanks to the fact that $A$ is a Toeplitz matrix \cite{toeplitz3}.

Analogous calculations for the second moment $m=2$ lead to
\begin{equation}
\frac{\partial ^\alpha}{\partial |x_0|^\alpha}\langle t^2 \rangle(x_0) =-2 \langle t^1 \rangle(x_0).
\end{equation}
More generally, the moments of the first passage time distribution are obtained recursively from
\begin{equation}
\frac{\partial ^\alpha}{\partial |x_0|^\alpha}\langle t^m \rangle(x_0) =-m \langle t^{m-1} \rangle(x_0),
\end{equation}
for $m=1, 2, \cdots $.

This above expression can be rewritten as
\begin{equation}
\left(\frac{\partial ^{\alpha}}{\partial |x_0|^{\alpha}}\right)^m \langle t^m \rangle(x_0) =(-1)^m \Gamma(m+1).
\end{equation}
Solving numerically this relation, namely $\langle t^m \rangle(x_0)=(-1)^m\Gamma(m+1)(L/M)^{m \alpha}A^{-m}{\bf 1}$, allows us to compute all the moments of the first passage times distribution, which is akin to knowing the full distribution.

\section{Conclusions}
\label{conclusions}

In this paper, we have studied the eigenvalue-eigenfunction problem for the fractional Laplacian of order $\alpha$ with absorbing and free BC on a bounded domain. This problem has applications to many physical systems, including L\'evy flights and stochastic interfaces. We have proposed a discretized version of the operator whose properties are better suited to  bounded domains. It does not suffer from any slowing down in convergence and can easily take into account BC. When $\alpha \le 2$, the discrete fractional Laplacian may be interpreted in the light of two physical models for hopping particles and for elastic springs, where the BC emerge naturally and are easily implemented. An analytical continuation for $\alpha>2$ is also discussed. Our approach easily allows to obtain the numerical eigenfunctions and eigenvalues for the fractional operator: eigenfunctions corresponding to absorbing BC show the expected power-law behavior at the boundaries. We also gain analytical insights into the problem by calculating perturbative corrections for the eigenvalues around $\alpha=0$ and $2$. Further information on the eigenvalue structure is obtained by studying the case of even $\alpha$, where a semi-analytical treatment is possible: for every $\alpha$ the spectra seem to approach exponentially fast a simple functional form. This conjecture has been proven for the case of even $\alpha$ and is supported by numerical investigations for real $\alpha$. The first passage problem and its connection to the fractional Laplacian operator were also explored.

\acknowledgments
This work was supported by the NSF grant DMR-04-2667 (M.K.). We are grateful for support from the Fondazione Fratelli Rocca through a Progetto Rocca fellowship (A.Z.), and from a Pierre Aigrain fellowship (A.R.).

\appendix
\section{Additional Notes}
\label{appendix}

\subsection{Integral representation of Riesz derivatives}
\label{riesz derivatives}
Riesz fractional derivatives are defined as a linear combination of left and right Riemann-Liouville derivatives of fractional order, namely
\begin{equation}
\frac{d^{\alpha}}{d|x|^{\alpha}} f(x)=-\frac{1}{2\cos((m-\alpha)\pi /2)} \left[ {\cal D}_+^\alpha - {\cal D}_-^\alpha \right],
\label{integral definition}
\end{equation}
where
\begin{equation}
{\cal D}_+^\alpha=\frac{1}{\Gamma(\alpha)} \int^{x}_{a} \dd y \left(x-y \right)^{m-\alpha-1} f^{(m)}(y)
\end{equation}
and
\begin{equation}
{\cal D}_-^\alpha=\frac{1}{\Gamma(\alpha)} \int^{b}_{x}  \dd y\left(y-x \right)^{m-\alpha-1} f^{(m)}(y) ,
\end{equation}
with $\alpha \in (m-1,m)$, $m$ integer, and $x \in \Omega=[a, b]$. This definition does not hold for odd $\alpha$. The integrals in Eq.~(\ref{integral definition}) have a power-law decaying kernel \cite{podlubny,samko}.

\subsection{Eigenfunctions of $-(-\triangle)^{\frac{\alpha}{2}}$ for even $\alpha$}
\label{regular laplacian}
When $\alpha=2$ the operator in Eq.~(\ref{problem}) is the regular Laplacian. For the case of absorbing BC we impose $\psi_k(-1)=\psi_k(1)=0$ and get
\begin{equation}
\psi_{k}(x) =
\gld
\begin{array}{cc}
\cos(\frac{k \pi x}{2}) & \text{ when $k$ is odd} \\
\sin(\frac{k \pi x}{2}) & \text{ when $k$ is even}
\end{array}
\right. .
\end{equation}
The associated eigenvalues are $\lambda_k= (k \pi/2)^2$, where $k=1,2,\cdots $. For the case of free BC we impose $\psi_k^{(1)}(-1)=\psi_k^{(1)}(1)=0$ and get
\begin{equation}
\psi_{k}(x) =
\gld
\begin{array}{cc}
\cos\left(\frac{(k-1) \pi x}{2}\right) & \text{ when $k$ is odd} \\
\sin\left(\frac{(k-1) \pi x}{2}\right) & \text{ when $k$ is even}
\end{array}
\right. .
\end{equation}
The associated eigenvalues are $\lambda_k= ((k-1) \pi/2)^2$, where $k=1,2,\cdots $. For mixed BC, namely $\psi_k(-1)=\psi_k^{(1)}(1)=0$, we have
\begin{multline}
\psi_{k}(x) = \pm \frac{1}{\sqrt{2}}\left( \cos\left(\frac{(2 k-1) \pi x}{4}\right) \right.
\\ \left.+(-1)^{k+1}  \sin\left(\frac{(2 k-1) \pi x}{4}\right) \right) .
\end{multline}
and the associated eigenvalues are $\lambda_k= ((2 k-1) \pi/4)^2$, where $k=1,2,\cdots $.

For  absorbing BC, we present here also the analytical expressions for the eigenfunctions corresponding to the first even values of $\alpha$. For $\alpha=4$, the condition $\det(B)=0$ becomes $\cos(2 \Lambda_k)\cosh(2 \Lambda_k)=1$, whose first roots are $\Lambda_1= 2.36502\cdots $, $\Lambda_2=3.9266\cdots $, and so on. Correspondingly, the normalized eigenfunctions are
\begin{equation}
\psi_{k}(x) =
\gld
\begin{array}{cc}
\frac{\cos(\Lambda_k x)}{\sqrt{2}\cos(\Lambda_k)} - \frac{\cosh(\Lambda_k x)}{\sqrt{2}\cosh(\Lambda_k)} & \text{  when $k$ is odd} \\
\frac{\sin(\Lambda_k x)}{\sqrt{2}\cos(\Lambda_k)} - \frac{\sinh(\Lambda_k x)}{\sqrt{2}\cosh(\Lambda_k)}  & \text{ when $k$ is even}
\end{array}
\right. .
\end{equation}

For the case $\alpha=6$, due to a highly symmetric structure of the determinant equation, eigenfunctions may be expressed in close form. For example, the normalized ground state eigenfunction is 
\begin{multline}
\psi_{1}(x)= \tanh\left(\frac{\sqrt{3}\pi}{2}\right) \cos(\pi x)\\
+\frac{\sqrt{3}}{\cosh(\sqrt{3} \pi /2)} \cos\left(\frac{\pi}{2}x\right) \cosh\left(\frac{\sqrt{3}\pi}{2}x\right)\\
+ \frac{1}{\cosh(\sqrt{3} \pi /2)} \sin\left(\frac{\pi}{2}x\right) \sinh\left(\frac{\sqrt{3}\pi}{2}x\right).
\end{multline}

\end{document}